\documentclass[preprint]{ptephy_v1}

\preprintnumber{XXXX-XXXX} 

\usepackage{subfigure}

\begin{document}
\title{Quantum Mechanical Approach to Bifurcation Point Detection in Hamiltonian Dynamical Systems.}
%
%
\author{Hironori Makino}
\affil{Department of Human and Information Science at Tokai University, Hiratsuka 259-1207, Japan\email{makino@tokai-u.jp}}
%
%
%

\begin{abstract}
Energy level statistics of a bounded quantum system, whose classical dynamical system exhibits bifurcations, is investigated using the two-point correlation function (TPCL), which at the bifurcation points exhibits periodic spike oscillations owing to the accumulation of levels called the shell effect.  The spike oscillations of the TPCL is analyzed by the reduced chi-squared value which deduced to exhibit abrupt increases at bifurcation points, thereby yielding a novel detection approach.  Using this method, we attempt to numerically detect the bifurcation points of a lemon-shaped billiard. 
\end{abstract}
\subjectindex{xxxx, xxx}
\maketitle
\section{Introduction}
Bifurcation point detection is a crucial technique in the research field of nonlinear dynamical systems, which has played a significant role in understanding the nonlinear phenomena in the actual world\cite{KUZ}. There are several conventional approaches based on the Newton-Raphson method\cite{NRM1,NRM2}, and more recently, the meta-heuristic approaches, such as the particle swarm optimization (PSO), the differential evolution (DE) and the evolution strategy (ES), have garnered considerable attention\cite{PSO1,PSO2,DE,ES}. Because these methods rely on an objective function obtained from the eigenvalues of the monodromy matrix (Jacobi matrix of variational equations) to evaluate the stability of each periodic point(fixed point), it is difficult to determine the bifurcation point when the matrix cannot be easily obtained. The derivation of the objective function is generally difficult in most nonlinear dynamical systems that cannot be solved analytically, and it is not easy to determine the position of the fixed points, let alone analyze its stability.  The objective of this study is to propose a novel approach to the bifurcation-point detection for Hamiltonian dynamical systems that have quantum mechanical counterparts.  Instead of the objective function, the bifurcation point is detected by the quantum mechanical data of the eigenenergy levels.  The core principle that makes this possible is a physical phenomenon called the shell effect \cite{Shell1,Shell2}.

The shell effect is a phenomenon in which the degeneracy of periodic orbits due to the bifurcation produces a periodic strong accumulation of eigenenergy levels in the corresponding quantum system. This phenomenon has been reported to exist in bounded quantum systems such as atomic nuclei, metallic clusters, and mesoscopic semiconductor systems such as quantum dots, and quantum billiards, as well as in partially open quantum systems; in addition the related research field is still  expanding\cite{SSIVS1,SSIVS2,SSIVS3,SSIVS4,SSIVS5,SSIVS6,SSIVS7,SSIVS8,SSIVS9,SSIVS10,SSIVS11}.  On the other hand, its mechanism has been elucidated by the semiclassical theory\cite{Scho,Mag,Mag2}, where the quantum effects of bifurcation in the statistical properties of energy levels have also been elucidated using an extended version of the Gutzwiller's trace formula that addressed the divergence problem at the bifurcation point\cite{Gutzw}.
Numerical attempts to verify the impact of the shell effect on energy level statistics have been made. Berry, Keating,  and Prado investigated the level number variance (LNV) of the perturbed cat map at a saddle-node (tangent) bifurcation and reported that there is an additional contribution to the long-range spectral correlation called the "lift-off," which causes the LNV to increase rapidly at a certain correlation length determined by the semiclassical theory\cite{BKP}. The lift-off was observed also for the pitchfork bifurcation of the coupled quartic oscillators by Gutiérrez et al.\cite{GBRS}. For the short-range spectral correlation, Makino, Harayama, and Aizawa numerically investigated the nearest-neighbor level-spacing distribution (NNLSD) of the quantum oval billiard, and then reported anomalous accumulation between adjacent levels at the bifurcation\cite{MHA99}.  These phenomena were recently investigated by Makino in terms of the two-point correlation function (TPCF), where the close relationship to the shell effect that causes periodic spike accumulation of levels is elucidated\cite{MAK}.

The TPCF is defined as the probability density of identifying two levels at a specific energy distance. In the research field of quantum chaology, this function is typically derived from the eigenenergy levels in a sufficiently small energy range $[E, E+\Delta E]$.  This is because the phase space structure of Hamiltonian system is generally energy-dependent, and fixing the dynamics for all eigenstates considered in the statistics provides a clear correspondence between the quantum and classical aspects. For the bifurcation detection method proposed in this paper to work well, this point must be taken into account for the energy range to be analyzed. We will discussed this point again in section Section \ref{sect6}.  Note that for the systems with $f$ degrees-of-freedom, the number of levels in the interval $\Delta E$ diverges as $O(\Delta E/\hbar^f)$ in the semiclassical limit $\hbar\to 0$, indicating that the small interval $\Delta E$ still contains a sufficient number of levels to ensure good statistical significance in the deep semiclassical regime.  The billiard systems have a convenient scaling law that makes the phase space structure invariant with respect to energy $E$, and hence, the entire energy spectrum can be used in the analysis.  In this paper we will analyze the one-parameter family of lemon-shaped billiard whose bifurcation points are well understood \cite{Heller,Reichl,MAK01,MAK18}, and will attempt to estimate the bifurcation points by the quantum mechanical data of eigenenergy levels characterized by the TPCF.

The remainder of this paper is organized as follows.  The lemon-shaped billiard is introduced in Section \ref{sect2}, while  bifurcation parameters are determined analytically in Section \ref{sect3}.  In Section \ref{sect4}, the TPCF of the quantum lemon billiard that exhibits periodic spike oscillations is analyzed.  These oscillations are evaluated by the Piason's $\chi^2$-test in Section \ref{sect5}, where the estimated values of the bifurcation parameters are determined by $\chi^2$ values obtained from the quantum mechanical data of eigenenergy levels.  The quantum mechanical estimates are compared with the bifurcation parameters of the classical dynamical system, followed by a summary and discussion in Section \ref{sect6}.

\section{Lemon-shaped billiard}
\label{sect2}
\begin{figure}[b]
\begin{center}
\includegraphics{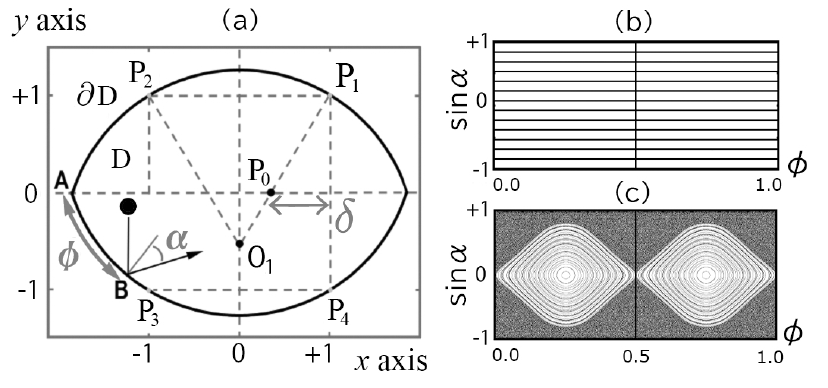}
\end{center}
\caption{(a):Definition of the billiard wall $\partial$D. We consider a square in the $x$-$y$ plane, whose four vertices $(\pm1,\pm1)$ are specified as $\mbox{P}_i,i=1-4$. Let $\mbox{P}_0$ be the point $(1-\delta,0)$ on the $x$-axis, where $\delta\in(0,1]$.  $\mbox{O}_1$ denotes the point where the extension of $\mbox{P}_1\mbox{P}_0$ intersects the y-axis, and it is the center of the circular arc $\mbox{P}_1\mbox{P}_2$. The wall $\partial$D is defined by the arcs $\mbox{P}_1\mbox{P}_2$ and $\mbox{P}_3\mbox{P}_4$ arranged symmetrically. (b)--(c): Poincar\'e surfaces of the section for (b)$\delta=1.0$ and  (c)$\delta=0.5$.}
\label{fig1}
\end{figure}
Figure \ref{fig1}(a) presents a schematic diagram of the lemon-shaped billiard whose boundary wall $\partial$D comprises two symmetrical arcs, where its curvatures are determined by a single parameter $\delta\in(0,1]$. For $\delta=1$(circular wall), the dynamical system is integrable, and the motion of a particle on the domain D is regular for any choice of initial conditions.  For $0<\delta<1$, the dynamical system is non-integrable and the motion of the particle on D is regular or chaotic, depending on the initial condition.

Figures \ref{fig1}(b) and 1(c) present the Poincar\'e surfaces of the section which, for the billiard problem, is described by the Birkhoff coordinates $(\phi,\sin{\alpha})$, where $\phi\in[0,1]$ represents the normalized curvilinear distance along the wall $\partial$D measured from the origin A to the collisional point B [refer to Fig.\ref{fig1}(a)], and $\alpha$ denotes the angle between the inner normal and the orbit reflected from the wall \cite{BKF}.  When $\delta=1$, the entire surface of the section is filled with the invariant tori[Fig.\ref{fig1}(b)].  As $\delta$ is altered from 1 to 0, the system transitions from the integrable to the non-integrable with successive bifurcations, while the surface of the section is filled with tori and chaos, as shown in Fig.\ref{fig1}(c).  The bifurcation parameters are derived in the next section.
\begin{figure}[t]
\begin{center}
\includegraphics{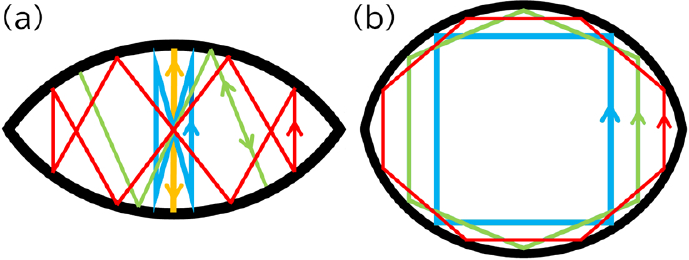}
\end{center}
\caption{Periodic orbits belonging to (a)the bouncing mode family and (b)the glancing mode family.}
\label{fig2}
\begin{center}
\includegraphics{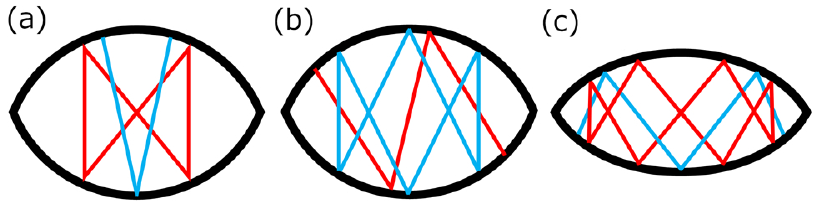}
\end{center}
\caption{Pairs of stable periodic orbit depicted in blue and unstable periodic orbit depicted in red, belonging to the bouncing mode family:(a)$n=2$,(b)$n=3$ and (c)$n=4$.}
\label{fig3}
\end{figure}
\section{BIFURCATIONS}
\label{sect3}
The periodic orbits observed in this system are a countability infinite set that can be divided roughly into two families: the bouncing mode family, as illustrated in the Fig.\ref{fig2}(a), comprising back-and-forth trajectories between  upper and lower walls, and the glancing mode family presented in Fig. \ref{fig2}(b), which comprises trajectories with repeated shallow reflections along the wall.  

The periodic orbits $n=2,3,4,\cdots$ belonging to the bouncing mode family form the pairs comprising
 a stable periodic orbit and an unstable periodic orbit of the same period $2n$, as illustrated in Figures \ref{fig3}(a)--(c), and each pair appears at the bifurcation point $\delta^{\mbox{\tiny B}}_n$ from the centers of the upper and lower arcs $\phi=0.5\pm0.25$ with an angle $\alpha=0$. The bifurcation parameters $\delta^{\mbox{\tiny B}}_n$ are determined by the equation $|\mbox{Tr}{M^n(\delta)}|=2$.  Here, $M(\delta)$ is the monodromy matrix obtained by linearizing the Poincar\'e map at $\phi=0.5\pm 0.25$ and $s\equiv\sin{\alpha}=0$ as(also refer to Ref.\cite{Berry81})
\begin{equation}
M(\delta)=\frac{\partial(\phi_{l+1},s_{l+1})}{\partial(\phi_l,s_l)}
=
\left[
\begin{array}{cc}
1-2w(\delta) & 2w(\delta)R(\delta) \\
-2[1-w(\delta)]/R(\delta) & 1-2w(\delta) \\
\end{array}
\right],
\end{equation}
\begin{figure}[b]
\begin{center}
\includegraphics{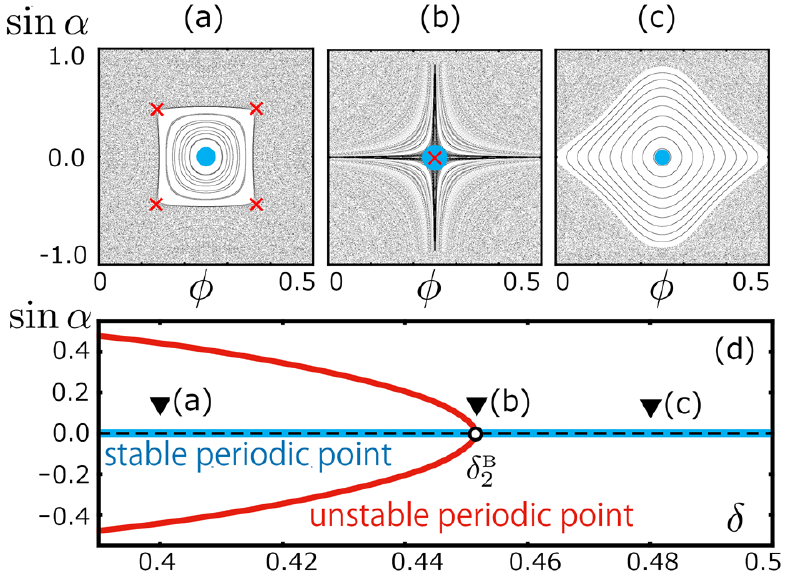}
\end{center}
\caption{(a)--(c):Poincar\'e surfaces of the section around the bifurcation point $\delta_2^{\tiny{B}}$;(a)$\delta=0.400$, (b)$\delta=0.451\simeq\delta_2^{\tiny{B}}$ and  (c)$\delta=0.480$, where the red and blue marks correspond to the unstable and stable points with period $4$, respectively. (d): Bifurcation diagram around  $\delta_2^{\tiny{B}}$ where the the red and blue lines correspond to the unstable(hyperbolic) and stable(elliptic) points with period 4, respectively.}
\label{fig4}
\begin{center}
\includegraphics{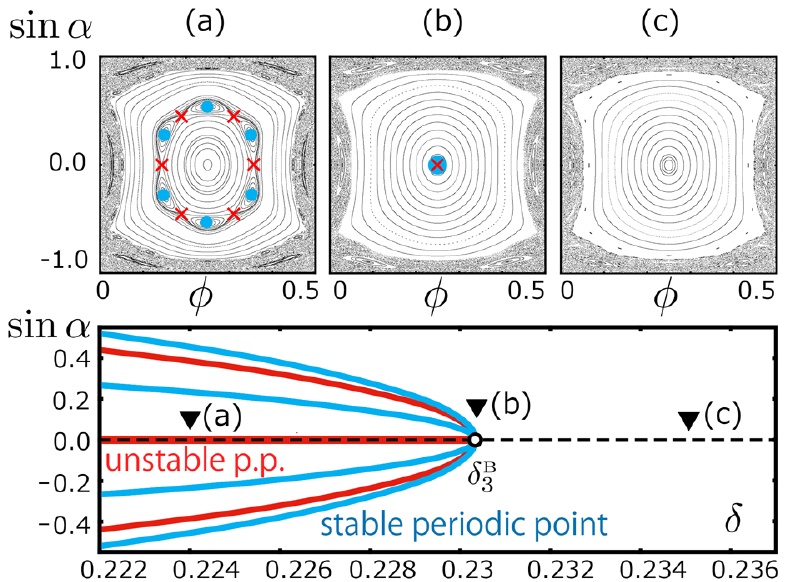}
\end{center}
\caption{(a)--(c):Poincar\'e surfaces of the section around the bifurcation point $\delta_3^{\tiny{B}}$;(a)$\delta=0.224$, (b)$\delta=0.230\simeq\delta_3^{\tiny{B}}$ and  (c)$\delta=0.235$, where the red and blue marks correspond to the unstable and stable points with period $6$, respectively. (d): Bifurcation diagram around $\delta_3^{\tiny{B}}$ where the the red and blue lines correspond to the unstable and stable points with period 6, respectively.}
\label{fig5}
\end{figure}
where $R(\delta)=\sqrt{1+\delta^2}/\delta$ represents the radius of circular arc $\partial D$, and $w(\delta)=1-(1-\delta)/\sqrt{1+\delta^2}$.  The solutions of the equation are obtained as $\delta^{\mbox{\tiny{B}}}_1=1$, $\delta^{\mbox{\tiny{B}}}_2=(4-\sqrt{7})/3$, $\delta^{\mbox{\tiny{B}}}_3=(16-3\sqrt{23})/7,\cdots$.  
In addition, the length of the bouncing periodic orbit $n$ at $\delta^{\mbox{\tiny{B}}}_n$ is
\begin{equation}
l^{\mbox{\tiny{B}}}_n = 4n\left( 1-\frac{1}{\delta^{\mbox{\tiny{B}}}_n} +R(\delta^{\mbox{\tiny{B}}}_n)\right).
\label{eq:lB}
\end{equation}
Figures \ref{fig4} and \ref{fig5} represent the Poincar\'e surfaces of the section and bifurcation diagram around the bifurcation points $\delta_2^{\tiny{B}}$ and $\delta_3^{\tiny{B}}$, respectively.  The red and blue marks or lines in each figure represent the unstable (hyperbolic) and stable (elliptic) periodic points, respectively, which merge at the bifurcation points. The bifurcation structures analyzed in this study are similar to the saddle-node type of one-dimensional systems.

Each periodic orbit with a period $2(n+1)$ belonging to the glancing mode family $n=1,2,3,\cdots$, emerges at 
$\delta_{n}^{\mbox{\tiny G}}\equiv \gamma_n - \sqrt{\gamma_n^2-1},\ \gamma_n=1+1/ \tan^2\alpha(n)$ 
from the positions $\phi=0$ and $0.5$, with an angle $\alpha(n)=n\pi/2(n+1)$, and exists in the parameter region $\delta\in[\delta_{n}^{\mbox{\tiny G}},1)$, 
where $\delta_{n}^{\mbox{\tiny G}}$ for $n=1-3$ are described as 
$\delta_{1}^{\mbox{\tiny G}}=2-\sqrt{3}$, $\delta_{2}^{\mbox{\tiny G}}=(4-\sqrt{7})/3$ and $\delta_{3}^{\mbox{\tiny G}}=4-2\sqrt{2}-\sqrt{23-16\sqrt{2}}$.  
The length of the glancing periodic orbit $n$ at $\delta_{n}^{\mbox{\tiny G}}$ is $l_n^{\mbox{\tiny{G}}}= 4nR(\delta_{n}^{\mbox{\tiny G}})\cos{\alpha(n)}$.

Table \ref{table:1-1} presents the parameters $\delta^{\mbox{\tiny B}}_n$ and $\delta^{\mbox{\tiny G}}_n$, including orbit lengths $l^{\mbox{\tiny B}}_n$ and $l^{\mbox{\tiny G}}_n$ for periodic orbits n = 1-9.  Because  $\delta_{n}^{\mbox{\tiny G}}$ has a property $\delta_{n}^{\mbox{\tiny G}}<\delta_{n+1}^{\mbox{\tiny G}}$ for all $n$, there are no periodic orbits belonging to the glancing-mode family in the region $\delta\in [0,\delta_1^{\mbox{\tiny G}})$.  Hence, we will focus on the region $\delta\leq0.5$ where the bifurcations of bouncing periodic orbits are mainly observed, and also explore the bifurcation points $\delta^{\mbox{\tiny B}}_n$ from the quantum mechanical data of eigenenergy levels.
\begin{table}[h]
  \caption{Bifurcation parameters of the bouncing and glancing periodic orbits, including the orbit length 
  at each bifurcation point.}
  \label{table:1-1}
  \centering
  \begin{tabular}{ccccc}
    \hline
 $n$ & $\delta_n^{\mbox{\tiny{B}}}$ &  $l_n^{\mbox{\tiny{B}}}$ & $\delta_n^{\mbox{\tiny{G}}}$  &  $l_n^{\mbox{\tiny{G}}}$  \\
    \hline \hline
   1& 1.000000 & 5.657 & 0.267949 & 10.93 \\
   2& 0.451416 & 9.722 & 0.451416 & 9.722\\
   3& 0.230358 & 13.36 & 0.561177 & 9.384\\    
   4& 0.138320 & 17.10 & 0.634095 & 9.233\\
   5& 0.0916968 & 20.92 & 0.686118 & 9.150\\
   6& 0.0650173 & 24.78 & 0.725133 & 9.097\\
   7& 0.0484028 & 28.68 & 0.755493 & 9.062\\
   8& 0.0373881 & 32.60 & 0.779796 & 9.035\\
   9& 0.0297253 & 36.53 & 0.799696 & 9.017\\
    \hline
  \end{tabular}
\end{table}
\section{Two-Point Correlation function}
\label{sect4}
The eigenenergy levels $E_\ell,\ell=1,2,3,\cdots$ of the quantum lemon billiard are obtained by solving the time-independent Schrödinger–Helmholtz equation $\nabla^2\varphi(\textbf{r})$+$E\varphi(\textbf{r})=0$ under the Dirichlet boundary condition $\varphi(\textbf{r}\in\partial\mbox{D})=0$, and are transformed to a stationary point process $\{\epsilon_\ell\}$ called {{\it unfolded}} energy levels, whose mean spacing is unity\cite{Boh}.  The transformation $\{E_\ell\}\to\{\epsilon_\ell\}$ is carried out by using the leading Weyl term of the integrated density of states, $\bar{N}(E)=AE/4\pi$, as $\epsilon_\ell=\bar{N}(E_\ell)$, where $A$ represents the area of the billiard domain D.  In a quantum lemon billiard that has four parity symmetry classes $\psi(\pm x,y)=\pm\psi(x,y)$ and $\psi( x,\pm y)=\pm\psi(x,y)$, the eigenenergy levels are divided into mutual independent components belonging to these four classes; hence, the {{\it unfolding}} transformation needs to be carried out separately as $\epsilon_\ell^{'}=\bar{N}(E_\ell^{'})/4$ for each of the four components, thereby yielding four unfolded sets of levels.  The TPCF analyzed in this study represents the probability density of identifying two levels at spacing $L$, and is defined for each of the energy level components using the level density $d(x)=\sum_\ell \delta(x-\epsilon_\ell^{'})$ as $R_2(L)=\left< d(x-L/2)d(x+L/2)\right>$, where the bracket $\left<\cdots\right>$ stands for an averaging over $x$\cite{Mehta}.  This quantity is suitable for studying the quantum-mechanical effects of bifurcation, as its relationship with classical periodic orbits is well understood in the semiclassical theory.  Based on Gutzwiller's trace formula\cite{Gutzw}, TPCF is expressed by the periodic-orbit sum as

\begin{eqnarray}
&&\left< d\left(\epsilon-\frac{L}{2}\right)d\left(\epsilon+\frac{L}{2}\right) \right>\simeq\sum_j | C_{j}(\epsilon) |^ 2 \cos{\left[\frac{T_j}{\hbar}L \right] } \nonumber\\
&&\qquad +O\left(\sum_{j_1\not=j_2}\left<\exp{\left[i\frac{S_{j_1}(\epsilon)-S_{j_2}(\epsilon) }{\hbar}\right] }\right>\right),\label{eq:3} \\\nonumber
\end{eqnarray}
\begin{figure}[t]
\begin{center}
\includegraphics{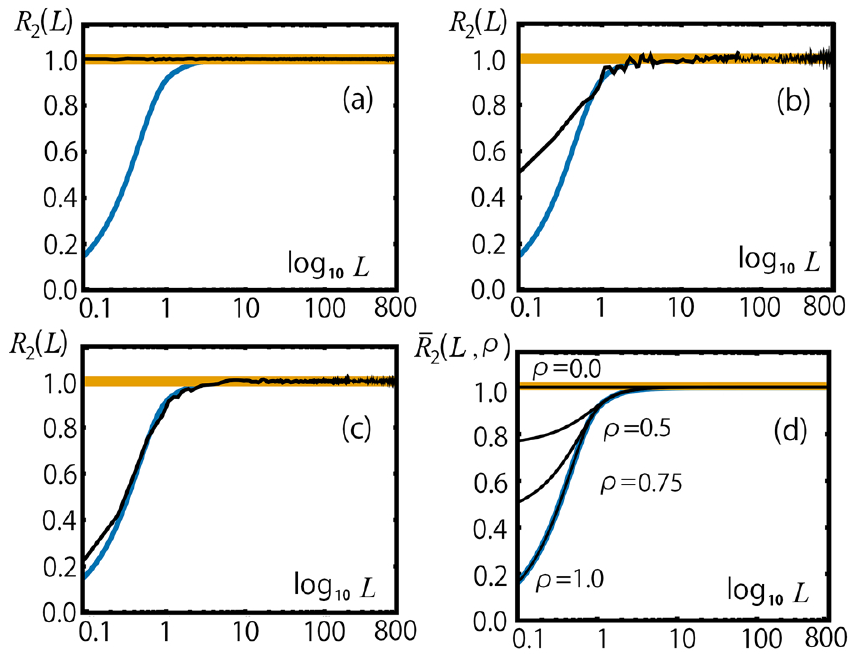}
\end{center}
\caption{(a)--(c): Numerical plots of TPCF $R_2(L)$ for (a)$\delta=1$, (b)$\delta=0.36$, (c)$\delta=0.43$, computed by (a)4840000 levels from $\epsilon=160000$ and (b)--(c) 8000 levels from $\epsilon=160000$. (d): Interpolation formula $\bar{R}_2(L;\rho)$ obtained for $\rho=0.0,0.5,0.75,$ and $1.0$.  The yellow curve at $\rho=0.0$ and blue curve at $\rho=1.0$ are $R_2^{\mbox{\tiny Poisson}}(L)$ and $R_2^{\mbox{\tiny{GOE}}}(L)$, respectively.}
\label{fig6}
\begin{center}
\includegraphics{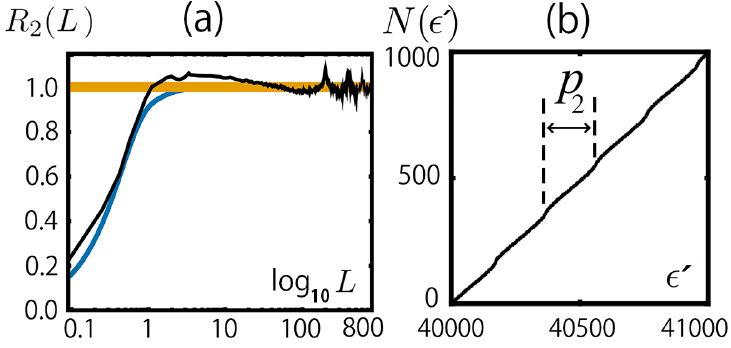}
\end{center}
\caption{(a):Numerical plots of TPCF $R_2(L)$ at the bifurcation point $\delta_2^{\mbox{\tiny B}}$, computed by 8000 levels from $\epsilon=160000$. (b): Integrated density of the odd-odd states at
$\delta_2^{\mbox{\tiny B}}$.}
\label{fig7}
\end{figure}
where $j$ labels each primitive periodic orbit and its repeatations, $C_j (\epsilon)$ represents the amplitude factor determined by the degeneracy and stability of the orbit, $S_j(\epsilon) $ denotes the action integral along the orbit $j$, which is defined here to include the Maslov index, and $T_j =\partial S_j(\epsilon)/\partial\epsilon$ represents the time period of the periodic orbit.  The second term in the RHS of Eq.(\ref{eq:3}) is expected to disappear via the smoothing procedure over $\epsilon$, TPCF is approximated by a sum of periodic functions,
\begin{equation}
R_2(L)\sim \sum_j| C_{j}(\epsilon) |^ 2 \cos{\left[\frac{T_j}{\hbar}L\right] }
\label{R2SC}
\end{equation}
whose periods with respect to $L$ are described as $p_j=2\pi\hbar/T_j$. For the billiard problem, this period is rewritten by the orbit length $l_j$ as
\begin{equation}
p_j=\frac{4\pi}{l_j}\sqrt{\epsilon}.
\end{equation}
It should be noted that the creation of a periodic orbit $j^*$ across the bifurcation generates an additional contribution 
$| C_{j^*}(\epsilon) |^ 2 \cos{\left[T_{j^*} L /\hbar\right] }$ to Eq.(\ref{R2SC}) and the creation of this term can exert a significant impact on the property of the TPCF, if the amplitude $C_{j^*}$ is relatively non-negligible in an infinite series $\sum_j$.  Such a possibility is expected to emerge at the bifurcation point where the extended trace formula, which is derived from the improved stationary-phase-approximation, predicts a significantly large and non-divergent value of $C_{j^*}$(also refer to Ref.\cite{Scho,Mag,Mag2}).  Furthermore, this should trigger periodic oscillations of the $p_{j^*}$ period in the behavior of the TPCF.

Figures \ref{fig6}(a)--\ref{fig6}(c) present the numerical plots of TPCF for various values of $\delta$, 
where the eigenenergy levels are obtained by the boundary element method, and the TPCF analyzed in this study is ultimately determined by the superposition of four TPCFs obtained respectively from the {\it unfolded} energy levels of the four parity-symmetry classes.

It is widely known that in time-reversal invariant quantum systems with classically fully chaotic counterpart, a universality proposed by Bohigas, Giannoni, and Schmit(BGS) exists\cite{BGS}, such that the {\it unfolded} energy levels in the semiclassical limit exhibit the same fluctuation properties as predicted by the gaussian orthogonal ensemble (GOE) statistics of the random matrix theory\cite{Mehta}, which provides the TPCF in the following form 
\begin{equation}
R_2^{\mbox{\tiny GOE}}(L)= 1-\sigma^2(L) -\frac{d\sigma(L)}{dL}\int_L^{+\infty}\sigma(L')dL',
\end{equation}
where $\sigma(L)=\sin(\pi L)/(\pi L)$[also refer to the blue curves in Figure.\ref{fig6}].  While in quantum systems with a classically integrable counterpart, another universality proposed by Berry and Tabor exists\cite{BT}, such that the {\it unfolded} energy levels in the semiclassical limit exhibit the same fluctuation properties as the random number from the Poisson point process, which gives $R_2^{\mbox{\tiny Poisson}}(L)= 1$[refer to the yellow lines of Figure \ref{fig6}].  The theoretical underpinnings of these two universalities remain a subject in the research field of quantum chaology\cite{Berry85,Muller04,Keating07,Mark,Eskin}.  For a quantum system whose classical dynamical system comprises regular and chaotic motions, the TPCF fits neither $R_2^{\mbox{\tiny Poisson}}(L)$ nor $R_2^{\mbox{\tiny{GOE}}}(L)$ as shown in Figs.\ref{fig6}(b) and \ref{fig6}(c). In this case, it is useful to introduce their interpolation formula[refer to Fig.\ref{fig6}(d)]
\begin{equation}
\bar{R}_2(L;\rho)=\rho R_2^{\mbox{\tiny GOE}}(\rho L)+(1-\rho)R_2^{\mbox{\tiny Poisson}}\left((1-\rho)L\right),
\label{R2BRF}
\end{equation}
whose physical meaning is supported by the Berry-Robnik level statistics\cite{Rob98,BR84}. In the Berry-Robnik level statistics, Eq.({\ref{R2BRF}}) provides the TPCF of eigenenergy levels, which is a product of the statistically independent superposition of two spectral components following the GOE and Poisson statistics, while the relative weight $\rho\in[0,1]$ is assumed to coincide with the relative phase volume in the Liouville measure of the chaotic component.  In this research, we do not go into its physical meaning and deal with $\rho$ as a fitting parameter.

Figure \ref{fig7} presents the TPCF $R_2(L)$ at the bifurcation point $\delta_2^{\mbox{\tiny B}}$ and the integrated density of parity odd-odd states $N(\epsilon')=\int_{40000}^{\epsilon'} dx\sum_{\ell} \delta(x-\epsilon_{\ell}^{\mbox{\tiny {odd-odd}}})$ where $\epsilon'=\epsilon/4$.  In this case, the TPCF exhibits strong correlations $R_2(L)>1$ at some intervals, and does not fit either $R_2^{\mbox{\tiny GOE}}(L)$, $R_2^{\mbox{\tiny Poisson}}(L)$, or their interpolation $\bar{R}_2(L;\rho)$ at all.  Note that this property emerges from the  periodic accumulation of levels with a certain period, as shown in Fig.\ref{fig7}(b).

\begin{figure}[t]
\begin{center}
\includegraphics{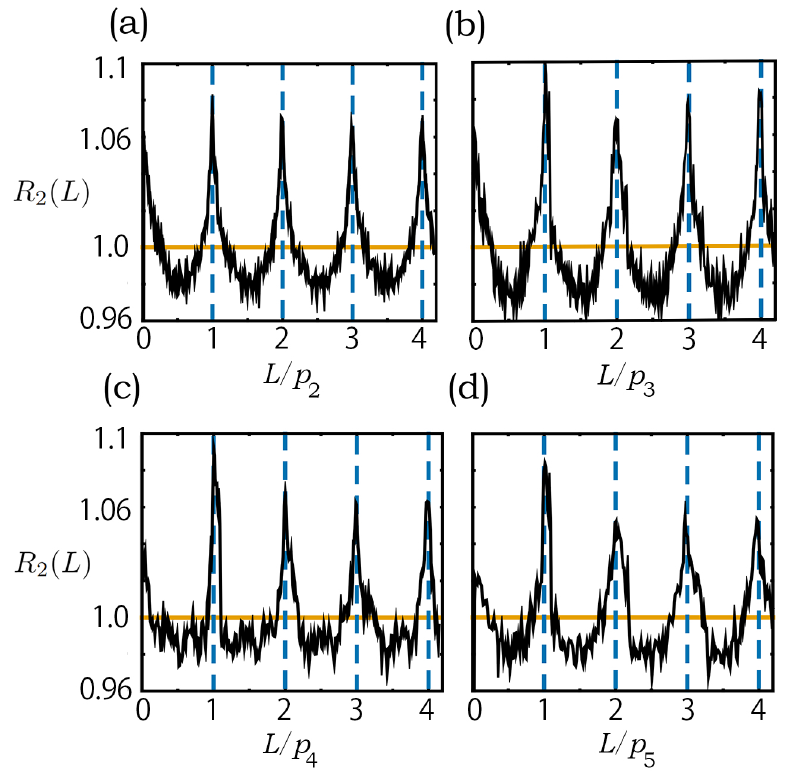}
\end{center}
\caption{Numerical plots of the TPCF $R_2(L)$ for (a)$\delta=\delta_2^{\mbox{\tiny B}}$, (b)$\delta=\delta_3^{\mbox{\tiny B}}$, (c)$\delta=\delta_4^{\mbox{\tiny B}}$ and (d)$\delta=\delta_5^{\mbox{\tiny B}}$, which are computed by $8000$  levels starting from $\epsilon=160000$.  The horizontal axis in each figure is rescaled by the period $p_n$ predicted by the semiclassical formula (\ref{R2SC}), whose value is determined by the median of the energy range, $\epsilon=164000$, as (a)$p_2\simeq 199.9$, (b)$p_3\simeq 161.1$, (c)$p_4\simeq 139.3$ and (d)$p_5\simeq 124.5$.}
\label{fig8}
\end{figure}

Figures \ref{fig8}(a)--(d) present the TPCF at the bifurcation points $\delta_n^{\mbox{\tiny B}},n=2-5$, respectively.  Here, the the horizontal axis in each figure is rescaled by the fundamental period $p_n=4\pi\sqrt{\epsilon}/l_n^{\mbox{\tiny B}}$, which is determined by the orbit length (\ref{eq:lB}) of the bifurcating orbit.  It is quite interesting that the TPCF in each figure exhibits remarkable spike oscillations whose period is well approximated by the fundamental period $p_n$ of the series $\sum_{j^*} | C_{j^*}(\epsilon) |^ 2 \cos{\left[T_{j^*} L/ \hbar \right] }$ in Eq.(\ref{R2SC}); hence, the oscillation is indeed contributed by the bifurcating periodic orbits $j^{*}\in\mbox{pair }n$.  In the next section, we propose an effective method to determine the bifurcation points from the quantum mechanical data of eigenenergy levels characterized by the TPCF.

\section{Bifurcation point detection}
\label{sect5}
As shown in the previous section, the bifurcation of the periodic orbit $j^* \in \mbox{pair }n$ triggers spike oscillations in the TPCF, which are $R_2(L)>1$ at $L=\ell p_n, \ell=0,1,2,\cdots$.  Consequently, $R_2(L)$ can no longer be approximated by conventional functions derived from the GOE statistics, the Poisson statistics, and their interpolation.  The idea proposed in this study is to adopt Pearson's $\chi^2$-test to effectively detect the occurrence of the spike oscillation.
\begin{figure}[t]
\begin{center}
\includegraphics{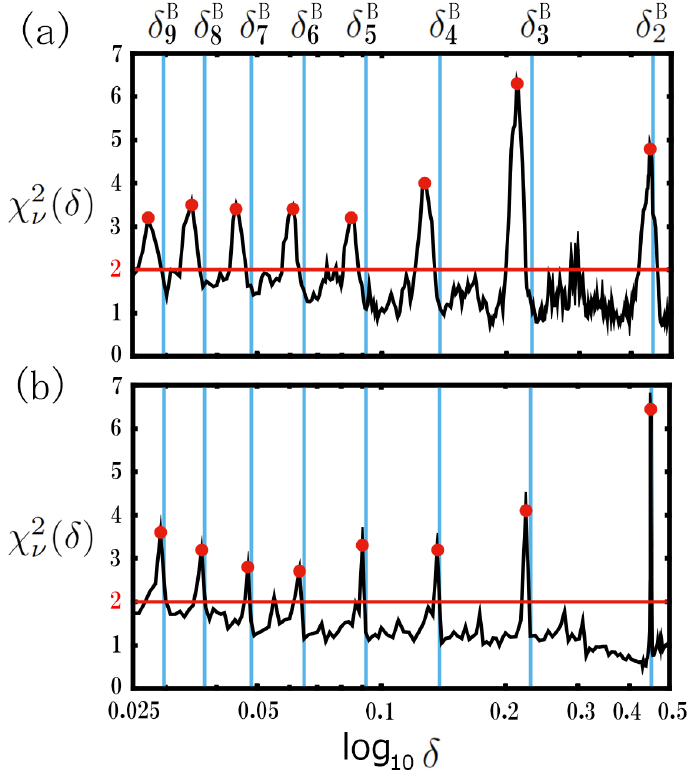}
\end{center}
\caption{Numerical plots of the reduced chi-squared value $\chi^2_\nu$ for 
various values of $\delta$, computed by (a)4000 levels from the ground states, (b)8000 levels from $\epsilon=160000$. }
\label{fig9}
\end{figure}

Pearson's $\chi^2$-test evaluates a measure $\chi^2$, which is a sum of differences between observed and expected outcome frequencies.  For a given TPCF $R_2(L)$, the observed frequency for each class $i=1,2,3,\cdots,K$ is calculated by using the class interval $\Delta L$ and total number $N$ of eigenenery levels as $o_i=N R_2(L_i)\Delta L$, while the expected frequency is calculated as $e_i =N \bar{R}_2(L_i,\rho)\Delta L$, where the interpolation parameter $\rho$ is determined by the best-fitting curve $\bar{R}_2(L,\rho)$ to the numerical data $R_2(L)$. Then, $\chi^2$ and its degrees of freedom $\nu$ are obtained as $\chi^2 = \sum_{i=1}^K (o_i-e_i)^2/e_i$ and $\nu=K-1$, where $\Delta L$ and $K$ are determined to hold $e_i>>1$ for every $i$, and determined to hold $e_i>5$ in our analysis.  If the reduced chi-squared value defined as $\chi^2_{\nu}\equiv\chi^2/\nu$ is less than 2, the conventional criteria for statistical significance, the goodness of the fitting is ascertained to be sufficient, and the null hypothesis that the numerical data $R_2(L)$ fits the formula $\bar{R}_2(L)$ is not rejected.  Conversely, in the case $\chi^2_\nu\geq2$, it is determined that $R_2(L)$ deviates significantly from $\bar{R}_2(L)$.  Such a case occurs when $R_2(L)$ exhibits a spike oscillation; hence, maximal points significantly  larger than 2 in the numerical plots of $\chi^2_{\nu}(\delta)$ provide candidates for the bifurcation points.

Figures {\ref{fig9}}(a)--(b) present the numerical plots of $\chi^2_\nu(\delta)$ in the parameter space, obtained for two different energy ranges.  The blue lines represent exact positions of the bifurcation points $\delta_n^{\mbox{\tiny B}}, n=2-9$ determined in the classical dynamical system.  Surprisingly, $\chi^2_\nu(\delta)$ in both energy ranges exhibits abrupt increases near the bifurcation points, each of which has its maximul value above the criterion of 2, as indicated by the red circle.  It should also be noted that $\chi^2_\nu(\delta)$ has its maximals closer to the bifurcation points in the higher energy region than in the lower energy region.

\begin{table}[tb]
\caption{Maxima of $\chi^2_\nu(\delta)$ in order from the largest value and their position $\delta_n^{\mbox{\tiny qm}}$, computed by (a)4000 levels from the ground state and (b)8000 levels from $\epsilon=160000$. The analytical solution of the bifurcation parameter $\delta_n^{\mbox{\tiny B}}$, which is closest to the position of each maximum, is also listed.}
\label{tab2}
\centering
\subtable[]{
\label{tab2a}
  \begin{tabular}{ccc}
    \hline
    $\chi^2_{\nu}(\delta_n^{\mbox{\tiny{qm}}})$ & $\delta_n^{\mbox{\tiny{qm}}}$  &  $\delta_n^{\mbox{\tiny B}}$  \\
    \hline \hline
    6.3  &  0.212$\pm$0.0014   & 0.230358 \\    
    4.8 &   0.444$\pm$0.0014   & 0.451416 \\
    4.0  &  0.127$\pm$0.0014   & 0.138320 \\
    3.5  &  0.0348$\pm$0.0008  & 0.0373881 \\
    3.4  &  0.0610$\pm$0.0008  & 0.0650173 \\
    3.4  &  0.0445$\pm$0.0008  & 0.0484028 \\
    3.2  &  0.0843$\pm$0.0008  & 0.0916968 \\
    3.2  &  0.0273$\pm$0.0008  & 0.0297253 \\
    2.5  &  0.297$\pm$0.0014   &  N/A \\
    2.2  &  0.293$\pm$0.0014   &  N/A \\
    2.2  &  0.287$\pm$0.0014   &  N/A \\
    2.1  &  0.0775$\pm$0.0008  &  N/A \\ 
    \hline
  \end{tabular}
}
\subtable[]{
\label{tab2b}
    \begin{tabular}{ccc}
    \hline
    $\chi^2_{\nu}(\delta_n^{\mbox{\tiny{qm}}})$ & $\delta_n^{\mbox{\tiny{qm}}}$  &  $\delta_n^{\mbox{\tiny B}}$  \\
    \hline \hline
   6.4  &0.451$\pm$0.001    &  0.451416 \\
   4.1  &0.224$\pm$0.001    &  0.230358 \\
   3.6  &0.0292$\pm$0.0003  &  0.0297253 \\
   3.3  &0.0900$\pm$0.0005  &  0.0916968 \\
   3.2  &0.137$\pm $0.001   &  0.138320 \\
   3.2  &0.0367$\pm$0.0003  &  0.0373881 \\
   2.8  &0.0475$\pm$0.0005  &  0.0484028 \\
   2.7  &0.0635$\pm$0.0005  &  0.0650173 \\
   2.2  &0.0552$\pm$0.0005  &  N/A\\
   2.0  &0.0867$\pm$0.0005  &  N/A \\
   1.9  &0.130$\pm$0.001    &  N/A \\
   1.8  &0.278$\pm$0.001    &  N/A \\
    \hline
  \end{tabular}
}
\end{table}

Tables \ref{tab2}(a)--(b) present the maxima of $\chi^2_\nu(\delta)$ for each of the two energy ranges, in order from the largest value, including their position $\delta_n^{\mbox{\tiny qm}}$ in the parameter space.  The analytical solution of the bifurcation parameter $\delta_n^{\mbox{\tiny B}}$, which is closest to the position of each maximum, is also listed in the Table.  In both energy ranges, the quantum mechanical data $\delta_n^{\mbox{\tiny{qm}}}$ obtained from large maximum 
value agree well with $\delta_n^{\mbox{\tiny B}}, n=2,3,4,\cdots$,
thereby providing an optimal estimate of the bifurcation points; in addition, the agreement is better in the higher energy region, which is in the deeper semiclassical regime.  However, for $\delta_n^{\mbox{\tiny{qm}}}$ obtained from the small maxima, there are no counterparts of $\delta_n^{\mbox{\tiny B}}$ that exhibit erroneous estimates. This result imply that the larger the $\chi^2_\nu(\delta_n^{\mbox{\tiny{qm}}})$ value, the better the estimate $\delta_n^{\mbox{\tiny{qm}}}$ is for the bifurcation parameter $\delta_n^{\mbox{\tiny B}}$. 
\section{Summary and discussion}
\label{sect6}
This study was dedicated to developing a novel method for obtaining estimates of the bifurcation points in Hamiltonian dynamical systems using the eigenenergy levels of its quantum systems characterized by TPCF.  Owing to the strong accumulation of levels caused by the shell effect, the TPCF at the bifurcation point exhibited periodic spike oscillations, whose period was well approximated by the semi-classical theory.  We introduced the Pearson's $\chi^2$ test to capture the spike oscillations of TPCF and determined that the reduced chi-squared value $\chi^2_\nu(\delta)$ exhibits abrupt increases at the bifurcation points, thereby providing an optimal estimate of the bifurcation parameters.  The accuracy of this estimation was determined to be better in the higher energy region, where the quantum system approaches the semiclassical domain.  

Note that the method proposed in this study is solely effective for Hamiltonian dynamical systems that have a quantum mechanical counterpart.  Because this method does not require the objective function or its derivative, it is very useful for analyzing systems whose bifurcation parameters are difficult to determine via the conventional analysis.

The proposed method described above has been applied to the lemon-shaped billiards with good results. And it could be effectively applied to more general Hamiltonian dynamical systems if the following condition is satisfied. 

The phase space geometry of Hamiltonian dynamical system is generally energy-dependent, and to obtain a clear correspondence between the quantum and classical aspects, the eigenenergy levels must be obtained from a sufficiently small interval $\Delta\epsilon$. However, when the interval is extremely small, our method may not detect bifurcations of short periodic orbits that produce level accumulations of long period.  In our method, the interval $\Delta\epsilon$ should contain at least one accumulation point, which provides the condition for the lower limit $\Delta\epsilon > p(\epsilon)$.  On the other hand, there is also the upper limit from another context.  The period of the spike oscillation $p(\epsilon)=4\pi\sqrt{\epsilon}/l$ is not constant over the range $[\epsilon,\epsilon+\Delta\epsilon]$, and the oscillation of $R_2(L)$ becomes incoherent at a position far from the origin.  Hence, it is also important to evaluate its coherence length $L_{\mbox{\tiny C}}$. Using the angular wavenumber $\omega(\epsilon)=2\pi/p(\epsilon)$ of the fundamental oscillation, it is defined by the phase shift $\Delta\omega(\epsilon)L_{\mbox{\tiny C}}=2\pi$ where $\Delta\omega(\epsilon)=\omega(\epsilon)-\omega(\epsilon+\Delta\epsilon)$.  The spike oscillations are maintained approximately in the region $L<L_{\mbox{\tiny C}}$, and our method works well when $L_{\mbox{\tiny C}}>> p(\epsilon+\Delta\epsilon)$.  This provides the condition $\Delta\epsilon << 3\epsilon$ and we have
\begin{equation}
\frac{4\pi}{l} \sqrt{\epsilon} <\Delta\epsilon<<3\epsilon.
\label{ineq7}
\end{equation}
For the unfolded energy levels  $\{\epsilon_\ell\}\in[\epsilon,\epsilon+\Delta\epsilon]$, our method is effective if the inequality (\ref{ineq7}) hold for the given interval $\Delta\epsilon$.  Numerical experiments for individual dynamical systems should be conducted in the future to collect specific examples and to determine the scope of effective application.  In the following, the above argument will be confirmed for the TPCF of the lemon-shaped billiard we analyzed.

\begin{figure}[h]
\begin{center}
\includegraphics{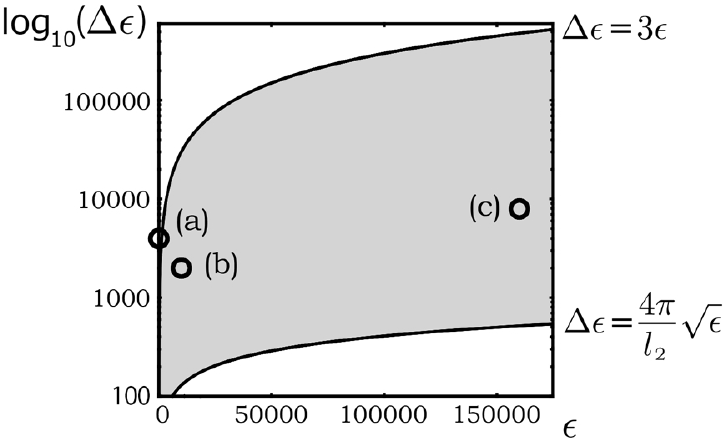}
\end{center}
\caption{Relationship between energy value $\epsilon$ and interval $\Delta\epsilon$ that satisfies the inequality (\ref{ineq7}). Plots (a)--(c) represent the energy ranges: (a) [0,4000], (b) [10000,12000] and (c) [160000,168000].}
\label{fig10}
\begin{center}
\includegraphics{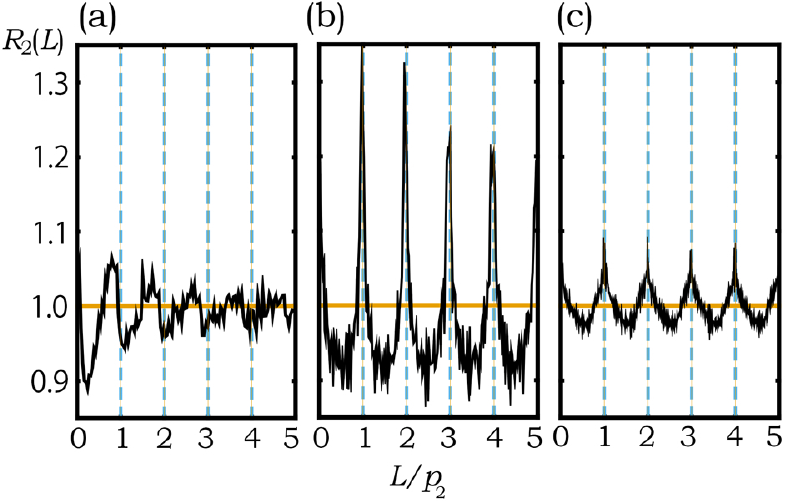}
\end{center}
\caption{Numerical plots of the TPCF $R_2(L)$ for $\delta=\delta_2^{\mbox{\tiny B}}$, which are computed by the eigenenergy levels supplied from the energy ranges: (a)[0,4000], (b)[10000,12000], and (c)[160000,168000].  The horizontal axis is rescaled by the fundamental period $p_2=4\pi\sqrt{\epsilon}/l_2^{\mbox{\tiny{B}}}\simeq 0.4911\sqrt{\epsilon}$ which is predicted by the semiclassical theory.}
\label{fig11}
\end{figure}

The gray region in Figure \ref{fig10} provides the energy range $[\epsilon,\epsilon+\Delta\epsilon]$ whose interval $\Delta\epsilon$ satisfies the inequality ({\ref{ineq7}}).  We also plotted the three energy ranges (a)[0,4000], (b)[10000,12000], and (c)[160000,168000], where the ranges (a) and (c) were adopted to identify the bifurcation points in Section {\ref{sect4}}.  Figures \ref{fig11}(a)--(c) present the TPCF at $\delta_2^{\mbox{\tiny{B}}}$ for the three energy ranges (a)--(c) of Fig. \ref{fig10}, respectively. The ranges (b) and (c) fully satisfy the condition ({\ref{ineq7}}), and the spikes of $R_2(L)$ are hence maintained over a region longer than the period $p_2(\epsilon)$, as shown in Figs.\ref{fig11}(b) and (c). While the range (a) has a short coherence length, and the spikes are collapsed within a range shorter than the period $p_2$ as shown in Fig.\ref{fig11}(a).

The reduced chi-square statistics is also applicable to the level-spacing distribution that is more popular in the research field of level statistics.  
In the present case, where some of the spectral components contributed from the bifurcating orbits, are strongly accumulated, it can be expected that the level-spacing distribution is consistent with a sub-Poisson or asymptotic Poisson distribution\cite{Mak03,Mak14}, and the reduced chi-square value is expected to exhibit abrupt increases in this statistics as well.  This possibility will be studied elsewhere.
\section*{Acknowledgment}
I thank Associate Professor Hiroshi Someya for helpful comments on recent developments in the optimization problem.


\begin{thebibliography}{9}
\bibitem{KUZ} Y.A. Kuznetsov, {\it Elements of Applied Bifurcation Theory} (Springer, 3rd Edition, 2004).
\bibitem{NRM1} H. Kawakami, {\it IEEE} Trans. Circuits Syst. {\bf 38}, 248 (1984).
\bibitem{NRM2} K. Tsumoto, T. Ueta, T.Yoshinaga and H.Kawakami, {\it IEICE} {\bf 3}, 458 (2012).
\bibitem{PSO1} J. Kennedy and R. Eberhart, Proceedings of the IEEE International Conference on Neural Networks, {\bf 4}, 1942 (1995).
\bibitem{PSO2} H. Matsushita, Y. Tomimura, H. Kurokawa and T. Kousaka, Int. J. Bifurcat. Chaos {\bf 27}, 1750101 (2017).
\bibitem{DE} R. Storn and K. Price,  J. Glob. Optim. {\bf 11}, 341 (1997).
\bibitem{ES} N. Hansen and A. Ostermeier, Proceedings of the IEEE International Conference on Evolutionary Computation, pp. 312-317 (1996)
\bibitem{GA} V. Chickarmane, S.R. Paladudgu, F. Bergmann and H.M. Sauro, Bioinformatics {\bf 18}, 3688 (2005).
\bibitem{Shell1} M.Brack and R.K.Bhaduri, {\bf Semiclassical Physics}(CRC Press, 2003).
\bibitem{Shell2} V. M. Strutinsky, Nucl. Phys. A {\bf 95}, 420 (1967); Nucl. Phys. A {\bf 122}, 1 (1968).
\bibitem{SSIVS1} R.Balian and C.Bloch, Ann. Phys. {\bf 69},76 (1972).
\bibitem{SSIVS2} V.M. Strutinsky, A.G. Magner, S.R. Ofengenden, and T. Døssing, Z.Phys. A {\bf 283}, 269 (1977).
\bibitem{SSIVS3} H. Frisk, Nucl. Phys. A {\bf 511}, 309 (1990).
\bibitem{SSIVS4} K. Arita and K. Matsuyanagi, Nucl. Phys. A {\bf 592}, 9 (1995).
\bibitem{SSIVS5} M. Brack, S. M. Reimann, and M. Sieber, Phys. Rev. Lett. {\bf 79}, 1817(1997). 
\bibitem{SSIVS6} A. Sugita, K. Arita and K. Matsuyanagi, Prog. Theor. Phys. {\bf 100} 597(1998).
\bibitem{SSIVS7} H. Nishioka, K. Hansen, and B. R. Mottelson, Phys. Rev. B {\bf 42}, 9377 (1990).
\bibitem{SSIVS8} M. Brack, J. Blaschke, S. C. Creagh, A. G. Magner, P. Meier, and S. M. Reimann, Z. Phys. D {\bf 40}, 276 (1997).
\bibitem{SSIVS9} S. M. Reimann, M. Persson, P. E. Lindelof, andM. Brack, Z. Phys. B {\bf 101}, 377 (1996).
\bibitem{SSIVS10} J. Blaschke and M. Brack, Europhys. Lett. {\bf 50}, 294 (2000).
\bibitem{SSIVS11} K. Arita and M. Brack, Phys. Rev. E {\bf 77}, 056211(2008).
\bibitem{Scho} H. Schomerus and M. Sieber, J. Phys. A {\bf 30}, 4537 (1997).
\bibitem{Mag} A.G. Magner, S.N. Fedotkin, K. Arita, T. Misu, K. Matsuyanagi, T. Schachner and M. Brack, Prog. Theor. Phys. {\bf 102},551(1999).
\bibitem{Mag2} A.G.Magner, S.N.Fedotkin, K.Arita, K.Matsuyanagi and M.Brack, Phys. Rev. E {\bf 63}, 065201(2001).
\bibitem{Gutzw} M.C. Gutzwiller, J. Math. Phys. {\bf 12},343(1971); M.C. Gutzwiller, {\it Chaos in Classical and Quantum Mechanics} (Springer, New York, 1990).
\bibitem{BKP} M.V.Berry, J.P.Keating and S.D.Prado, J. Phys. A {\bf 31}, L245(1998).
\bibitem{GBRS} M. Gutiérrez, M. Brack, K. Richter and A. Sugita, J. Phys. A {\bf 40}, 1525 (2007).
\bibitem{MHA99} H. Makino, T. Harayama and Y. Aizawa, Phys. Rev. E {\bf 59}, 4026 (1999).
\bibitem{MAK} H. Makino, Prog. Theor. Exp. Phys. 2019, 083A01 (2019).
\bibitem{Heller} E.J.Heller and S.Tomsovic, Phys.Today {\bf 46},38(1993).
\bibitem{Reichl} Suhan Ree and L. E. Reichl, Phys. Rev. E {\bf 60}, 1607 (1999).
\bibitem{MAK01} H. Makino, T. Harayama and Y. Aizawa, Phys. Rev. E {\bf 63}, 056203(2001).
\bibitem{MAK18} H. Makino, Prog. Theor. Exp. Phys. 2018, 073A01 (2018).
\bibitem{BKF} G.D. Birkhoff,{\it Dynamical Systems} (American Mathematical Society, Providence, RI, 1927; reprinted 1996).
\bibitem{Berry81} M.B.Berry, Eur. J. Phys. {\bf 2}, 91(1981).
\bibitem{Boh} O. Bohigas, {\it Random Matrices and Chaotic Dynamics}(North-Holland, Amsterdam, 1989) Les Houches Session LII. 
\bibitem{Mehta} M.L. Mehta, {\bf Random Matrices}(Elsevier, San Diego, CA, 2004), 3rd ed.
\bibitem{BGS} O. Bohigas, M.J. Giannoni and C. Schmit, Phys. Rev. Lett. {\bf 52}, 1 (1984).
\bibitem{BT} M.V. Berry and M. Tabor, Proc. R. Soc. Lond. A {\bf 356}, 375 (1977).
\bibitem{Berry85} M.V. Berry, Proc. Roy. Soc. Lond. A {\bf 400},229(1985).
\bibitem{Muller04} S. Müller, S. Heusler, P. Braun, F. Haake and A. Altland, Phys. Rev. Lett. {\bf 93}, 014103(2004).
\bibitem{Keating07} J.P. Keating and S. Müller, Proc. Roy. Soc. Lond. A {\bf 463}, 3241 (2007).
\bibitem{Mark} J. Marklof, Commun. Math. Phys. {\bf 199}, 169 (1998); Prog. Math.{\bf 202}, 421(2001); Duke Math. J. {\bf 115}, 409 (2002); Ann. Math.{\bf 158}, 419(2003).
\bibitem{Eskin} A. Eskin, G.A. Margulis and S.Mozes, Ann. Math.{\bf 161}, 679 (2005).
\bibitem{Rob98} M. Robnik, Nonlinear Phenom. Complex Syst. {\bf 1}, 1 (1998).
\bibitem{BR84} M.V. Berry and M. Robnik, J. Phys. A{\bf 17}, 2413 (1984).
\bibitem{Mak03} H. Makino and S. Tasak, Phys. Rev. E {\bf 67}, 066205 (2003).
\bibitem{Mak14} H. Makino and N. Minami, Prog. Theor. Exp. Phys. 2014, 073A01 (2014).
\end{thebibliography}
\end{document}